# The Minimum Mass of Planets, Dwarf Planets, and Planetary-scale Satellites

David G. Russell[a]

**Abstract**

The International Astronomical Union definitions for Planet and Dwarf Planet both require that a body has sufficient mass to overcome rigid body forces and self-gravitate into a nearly round shape. However, quantitative standards for determining when a body is sufficiently round have been lacking. Previously published triaxial ellipsoid solutions for asteroids, satellites, and Dwarf Planets in the radius range 135 – 800 km are examined to identify a minimum mass above which the entire population, regardless of composition, is round. From this data, the minimum mass to meet the roundness criterion is ≈5.0 x $10^{20}$ kg. The triaxial shape data suggests three radius ranges: (1) bodies with a radius < 160 km are non-spheroidal, (2) bodies with a radius in the range 160 – 450 km are transitional in shape or "nearly round", (3) bodies with a radius >450 km are spheroidal. Bodies orbiting the Sun with a mass greater than 5.0 x $10^{20}$ kg are Planets or Dwarf Planets. Arguments are presented for including the 16 spheroidal moons of the Solar System as a third dynamical class that can be identified as "Satellite Planets". Definitions are proposed that expand upon the taxonomy started in 2006 with the IAU Planet and Dwarf Planet classes.

~a: Author email: dgrussellastro@gmail.com
      Spencer, NY USA

**Introduction**

In 2006, the International Astronomical Union (IAU) approved definitions for classifying Solar System bodies orbiting the Sun as **Planets**, **Dwarf Planets**, or **Small Solar System Bodies**. In order to identify the correct class for a Solar System body orbiting the Sun two questions must be answered: (1) Is the body massive enough to self-gravitate into a nearly round shape? (2) Is the body dynamically dominant in its orbit?
   According to these definitions bodies with insufficient mass to self-gravitate into a spheroid are collectively identified as "*Small Solar System Bodies*" and include the various classes of asteroids, comets, Kuiper Belt objects and other trans-Neptunian bodies. Bodies with sufficient mass to self-gravitate into a spheroid are "*Planets*", if they are dynamically dominant in their orbit, or "*Dwarf Planets*", if they are not dynamically dominant in their orbit. The IAU definitions use the language *"has(has not) cleared the neighborhood around its orbit"* to indicate whether or not a body is dynamically dominant.
   The Planet and Dwarf Planet definitions were initially problematic because the "orbital clearing" and "roundness" criteria both lacked specific quantitative standards. In addition, specifying that Planets and Dwarf Planets "orbit the Sun" excludes exoplanets (Margot et al. 2024). Modifying the language to specify that a Planet orbits one or more stars, brown dwarfs, or stellar remnants allows the definition to be applied to exoplanetary systems (Margot 2015; Lecavelier des Etangs & Lissauer 2022; Margot et al. 2024).
   A metric for quantitatively determining whether or not a body has the minimum mass necessary to dynamically dominate or "clear its orbital zone" around a star, within a specified timescale, has been presented by Margot (Margot 2015 ; Margot et al. 2024, Equation 8). The minimum mass necessary to be dynamically dominant is not a single value, but instead increases with increasing orbital semi-major axis and increasing stellar mass (Margot et al. 2024, Figure 7). This metric can be used to determine whether or not a planetary body meets the standard of dynamical dominance in the Solar System and exoplanetary systems.

The roundness criterion is often considered problematic for the development of quantitative standards for a number of reasons. There are no agreed upon criteria for how round is round enough. In triaxial ellipsoid solutions, bodies that are oblate spheroids, such as the Earth, have identical a and b axis values with a shorter length c axis. However, there can be extreme cases such as the Dwarf Planet Haumea, in which a very short rotation period has led to three different triaxial ellipsoid axis lengths resulting in an oval spheroid shape (Ortiz et al. 2017).

The minimum mass and radius necessary to achieve a spheroidal shape also varies with the composition of the body. Rock composition bodies require a larger mass and radius to achieve a round shape than icy bodies (Lineweaver & Norman 2010). Given these difficulties, and whereas all currently known Solar System Planets and exoplanets have sufficient mass to self-gravitate into a spheroid, it has been argued that quantitative criteria for roundness are an unnecessary component of the "Planet" definition for the foreseeable future (Margot et al. 2024).

However, there are two important reasons why quantitative standards for determining whether or not a body is massive enough to be "round" are needed. First, it is possible for bodies with short period orbits around low mass stars and brown dwarfs to have insufficient mass to be spheroidal and still be dynamically dominant in their orbit (Soter 2006; Margot et al. 2024). In order to revise the IAU Planet and Dwarf Planet definitions to versions that can be consistently applied to all potential exoplanetary circumstances, a standard for the minimum mass necessary for a body to self-gravitate into a spheroid is needed. This minimum mass should be the same value for extrasolar bodies as the value used for Solar System bodies (Lecavelier des Etangs & Lissauer 2022). Second, since Dwarf Planets are not dynamically dominant in their orbits, the only criterion that can be used to distinguish a Dwarf Planet from a Small Solar System Body is the roundness criterion. Therefore, without a standard for the minimum mass necessary to be considered round, it is not possible to determine where the Dwarf Planet class ends and the Small Solar System Body class begins.

A number of studies have suggested possible minimum mass or radius values for attaining a round shape and qualifying as a Dwarf Planet. Stern & Levison (2002) suggested the minimum mass necessary for a body to be shaped primarily by gravity in less than a Hubble time is $\approx 10^{21}$ kg. For icy Dwarf Planets a minimum radius of ~225 km has been suggested by Tancredi (Tancredi & Favre 2008; Tancredi 2010). Lineweaver & Norman (2010) identified two minimum radii for Dwarf Planets based upon composition. For bodies with icy and rocky compositions, minimum radii for roundness of ~200 km and ~300 km respectively were indicated (Lineweaver & Norman 2010). The suggested minimum radius for icy bodies correspond to a minimum mass in the range ~3 - 6 x $10^{19}$ kg. Given that the asteroids Pallas and Vesta are not round, the minimum mass a rocky body needs to be round must exceed 3 x $10^{20}$ kg. Margot et al. (2024) noted that a mass of ~$10^{21}$ kg could be used as a minimum mass for determining if a body is approximately in hydrostatic equilibrium.

In this paper, previously published triaxial ellipsoid shape data for asteroids, icy moons, and Dwarf Planets is examined to evaluate how the "roundness" of bodies varies with mass and radius. Using b/a and c/a axis ratios, a single lower mass limit, independent of composition, is identified for Planets and Dwarf Planets. This lower mass limit can be applied to Solar System and exoplanetary bodies as a consistent mass boundary dividing Planets and Dwarf Planets from Small Solar (Stellar) System Bodies. In addition, the lower mass limit can serve as the minimum mass for planetary-scale satellites or "Satellite Planets".

This paper is organized as follows: In section 2 the minimum mass for Planets and Dwarf Planets is identified from previously published triaxial shape data. Section 3 provides arguments for extending the IAU Taxonomy to include planetary-scale satellites as a third dynamical class. In section 4, recommendations for extending and revising the taxonomy started with the 2006 IAU Planet and Dwarf Planet definitions are presented. Section 5 is the conclusion.

## ~2 Triaxial Ellipsoid Shape Data and the Minimum Mass of Planets and Dwarf Planets

### ~2.1 Triaxial Ellipsoid Shape Data from the Literature

In order to identify a useful minimum mass limit for self-gravitated "round" bodies, the literature was searched for published triaxial ellipsoid solutions for asteroids, icy moons, and dwarf planets in the radius range 135 – 800 km (Table 1). For each body with triaxial ellipsoid solutions, b/a and c/a axis ratios are calculated from the a, b, c axis values in the literature (Table 1). The bulk of the data comes from three sources: data for asteroids is from Vernazza et al. (2021), for the Saturnian satellites Roatsch et al. (2009), and for the Uranian satellites Thomas (1988). The data for Charon comes from Nimmo et al. (2017) and Proteus data is from Stooke (1994). Schaefer et al. (2008) did not provide triaxial data for Nereid, but determined the moon is non-spheroidal. Finally, triaxial data was not provided for Umbriel, Oberon, and Titania but all three bodies were determined to be spheroidal by Thomas (1988) and therefore the b/a axis ratio is assumed to be 1.000.

### ~2.2 Minimum mass for roundness indicated from shape data and axis ratios

Two difficulties with identifying a minimum mass for roundness have been that (1) the minimum mass appears to be different for rocky bodies than for icy bodies (Lineweaver & Norman 2010), and (2) there is not an objective standard for how round is round enough. In order to be applicable to exoplanetary systems, where composition can be uncertain, it is preferable to have a single minimum mass standard for all compositions rather than different standards for rocky and icy bodies.

Since the rotation of a body in a hydrostatic equilibrium state affects the shape of the body, generally resulting in an oblate spheroid shape rather than a perfect sphere, the b/a axis ratio is most important for determining whether or not each body is round. Bodies approach an increasingly "round" oblate spheroid shape as the b/a axis ratio approaches 1.000. Rather than a priori prescribing minimum b/a and c/a axis ratio values necessary to be considered sufficiently round, the axis ratios in Table 1 were examined to see how the b/a and c/a ratios change with increasing mass and radius. The focus was on identifying a minimum mass above which the ***entire sample*** of the Solar System's objects with triaxial shape data can be considered round regardless of composition.

The b/a and c/a axis ratios in Table 1 indicate the sample can be broken into three radius ranges (Figure 1). Bodies with radius <160 km have non-spheroidal shapes with b/a axis ratios <0.900 and c/a axis ratios <0.710. The non-spheroidal bodies all have a mass less than $3 \times 10^{19}$ kg. The second radius range is 160 – 270 km and can be considered "transitional" or "almost round". These bodies have b/a axis ratios in the range 0.920 – 0.980 and c/a axis ratios <0.970. Several bodies in this group (Mimas, Miranda, and Enceladus) have traditionally been included in the spheroidal group (Lineweaver & Norman 2010; Runyon et al. 2017; Margot et al. 2024). The mass range for this transitional group is $3 \times 10^{19}$ - $3 \times 10^{20}$ kg. The third radius range, bodies with radius > 450 km, is the spheroidal group. The bodies in this group all have b/a axis ratios > 0.980 and c/a axis ratios > 0.920. The lowest mass body in the spheroidal group is Tethys ($6.2 \times 10^{20}$ kg) and the smallest radius body is Ceres (469.7 km). It is also important to note that there is an approximately 200 km radius gap between the largest radius body in the "almost round" group and the smallest radius body in the spheroidal group. This gap provides a reasonable range within which the division between bodies that are round and those that are not round should exist.

**Table 1: Dwarf Planet, Asteroid, and Satellite shape data**

| Object | Mass (kg) | Radius (km) | a x b x c (km) | b/a | c/a | Shape Data Reference |
|---|---|---|---|---|---|---|
| **15 Eunomia** | $3.05 \times 10^{19}$ | 135 | 170 x 124 x 114.5 | 0.729 | 0.674 | Vernazza et al. (2021) |
| **Hyperion** | $5.55 \times 10^{18}$ | 135 | 180.1 x 133.0 x 102.7 | 0.738 | 0.570 | Thomas (2010) |
| **87 Sylvia** | $1.43 \times 10^{19}$ | 137 | 181.5 x 120 x 95.5 | 0.686 | 0.526 | Vernazza et al. (2021) |
| **511 Davida** | $2.66 \times 10^{19}$ | 149 | 179.5 x 146.5 x 126.5 | 0.816 | 0.705 | Vernazza et al. (2021) |
| **52 Europa** | $2.39 \times 10^{19}$ | 159.5 | 189 x 168 x 127.5 | 0.889 | 0.675 | Vernazza et al. (2021) |
| **704 Interamnia** | $3.52 \times 10^{19}$ | 166 | 177 x 171.5 x 151.5 | 0.969 | 0.856 | Vernazza et al. (2021) |
| **Nereid** | $3.1 \times 10^{19}$ | 179 | Non-spheroidal | | | Schaefer et al. (2008) |
| **Mimas** | $3.75 \times 10^{19}$ | 198.2 | 207.8 x 196.7 x 190.6 | 0.947 | 0.917 | Roatsch et al. (2009) |
| **Proteus** | $4.4 \times 10^{19}$ | 210 | 212 x 195 x 198 | 0.920 | 0.934 | Stooke (1994) |
| **10 Hygeia** | $8.74 \times 10^{19}$ | 216.5 | 225 x 125 x 212 | 0.956 | 0.942 | Vernazza et al. (2021) |
| **Miranda** | $6.29 \times 10^{19}$ | 235.8 | 240.4 x 234.2 x 232.9 | 0.974 | 0.969 | Thomas (1988) |
| **Enceladus** | $1.08 \times 10^{20}$ | 252.1 | 256.6 x 251.4 x 248.3 | 0.980 | 0.968 | Roatsch et al. (2009) |
| **2 Pallas** | $2.04 \times 10^{20}$ | 255.5 | 284 x 265 x 225 | 0.933 | 0.792 | Vernazza et al. (2021) |
| **4 Vesta** | $2.59 \times 10^{20}$ | 261.6 | 286.3 x 278.6 x 223.2 | 0.973 | 0.780 | Vernazza et al. (2021) |
| **1 Ceres** | $9.38 \times 10^{20}$ | 469.7 | 482.2 x 482.1 x 445.9 | 1.000 | 0.925 | Vernazza et al. (2021) |
| **Tethys** | $6.175 \times 10^{20}$ | 531.1 | 538.4 x 528.7 x 526.3 | 0.982 | 0.978 | Roatsch et al. (2009) |
| **Dione** | $1.10 \times 10^{21}$ | 561.4 | 563.4 x 561.3 x 559.6 | 0.996 | 0.993 | Roatsch et al. (2009) |
| **Ariel** | $1.23 \times 10^{21}$ | 578.9 | 581.1 x 577.9 x 577.7 | 0.994 | 0.994 | Thomas (1988) |
| **Umbriel** | $1.29 \times 10^{21}$ | 584.7 | Spheroidal | 1.000 | | Thomas (1988) |
| **Charon** | $1.59 \times 10^{21}$ | 606.0 | 606.5 x 606.5 x 607.0 | 1.000 | 1.001 | Nimmo et al. (2017) |
| **Iapetus** | $1.81 \times 10^{21}$ | 734.5 | 746.0 x 746.0 x 712 | 1.000 | 0.954 | Roatsch et al. (2009) |
| **Oberon** | $3.11 \times 10^{21}$ | 761.4 | Spheroidal | 1.000 | | Thomas (1988) |
| **Rhea** | $2.31 \times 10^{21}$ | 763.8 | 766.2 x 762.8 x 762.4 | 0.996 | 0.995 | Roatsch et al. (2009) |
| **Titania** | $3.46 \times 10^{21}$ | 788.9 | Spheroidal | 1.000 | | Thomas (1988) |

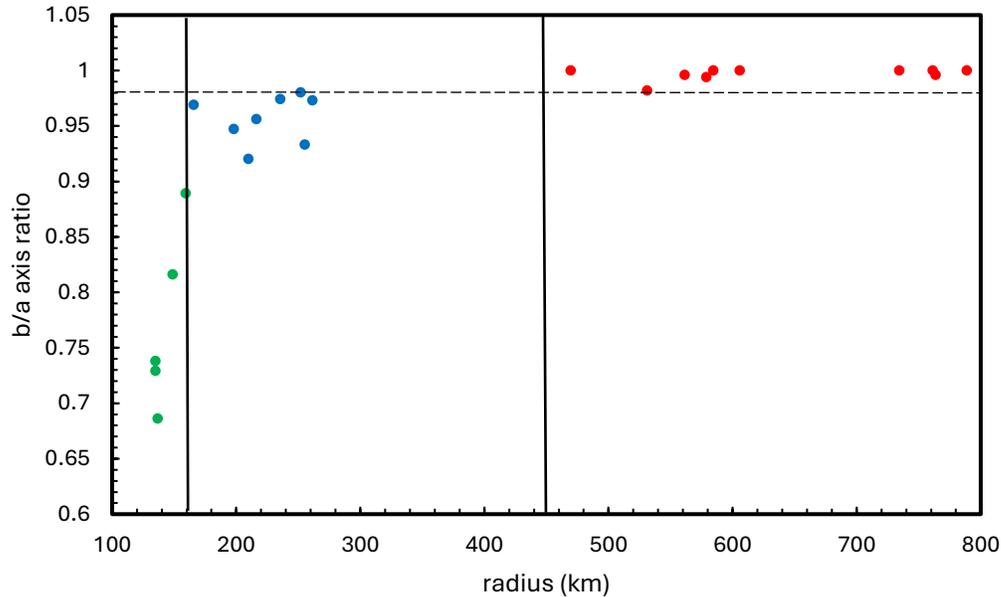

**Figure 1** – b/a axis ratio vs. radius for bodies in Table 1. Green dots are non-spheroidal bodies with b/a < 0.900. Blue dots are transitional bodies with b/a 0.900 – 0.980. Red dots are spheroidal bodies with b/a >0.980. Horizontal dashed line is for b/a axis ratio = 0.980.

From the data in Table 1, the sample characteristics identified for bodies meeting the Planet and Dwarf Planet roundness criterion are:

~ mass > 5 x $10^{20}$ kg
~ radius >450 km
~ b/a axis ratio > 0.980

Based upon these characteristics, **the minimum mass necessary to meet the roundness criterion for Planets and Dwarf Planets is ≈5 x $10^{20}$ kg.** Bodies orbiting the Sun with a mass below this mass limit are Small Solar System Bodies. Bodies exceeding this mass limit that orbit a star, brown dwarf, or stellar remnant are Planets or Dwarf Planets. Note that the b/a axis ratio of >0.980 should be treated as *generally descriptive* of bodies that are round, rather than a prescriptive minimum value for self-gravitated round bodies. The Dwarf Planet Haumea has a mass of 4 x $10^{21}$ kg (Ragozzine & Brown 2009) and radius of ~780 km (Dunham et al. 2019). Haumea's mass and radius values clearly exceed the minimum mass and radius standards for "roundness" observed in Table 1. However, due to an unusually fast rotation period of 3.9 hours, Haumea is not an oblate spheroid, but is instead an oval shaped triaxial ellipsoid (Ortiz et al. 2017) with a b/a axis ratio of ~0.73.

The icy moons Mimas, Miranda, and Enceladus have normally been included in the list of "round" Solar System bodies (Lineweaver & Norman 2010; Runyon et al. 2017; Margot et al. 2024). However, based upon the b/a axis ratios, all three moons can be classified as transitional, or "almost round" with b/a axis ratios in the range 0.947 to 0.980. In comparison, all of the spheroidal bodies, except Tethys, have b/a axis ratios of at least 0.994. Tethys, b/a = 0.982, is the lowest mass body in the spheroidal group with a mass of 6.18 x $10^{20}$ kg. With a rock core mass fraction of only ~6% (Thomas et al. 2007), Tethys can serve as a benchmark for the minimum mass a body must have to attain a spheroidal shape. In order for a bodies with high rock fractions to achieve a spheroidal shape, a mass greater than Tethys is necessary.  This is supported by the shape data for the three largest mass rocky asteroids Pallas, Vesta, and Ceres.  Pallas and Vesta have radii of ~260 km, masses less than $3 \times 10^{20}$ kg, and transitional b/a axis ratios of 0.933 and 0.973 respectively.   While the b/a axis ratios are approaching the minimum value for a round shape, the c/a axis ratios are < 0.800 for both asteroids demonstrating their transitional nature.  In contrast Ceres, with a smaller radius but 52% larger mass than Tethys, has b/a and c/a axis ratios of 1.000 and 0.925 respectively, indicating a spheroidal shape.

While candidate Trans-Neptunian Dwarf Planets lack triaxial solutions derived from close flyby and orbital mission data, the mass and radius values for these bodies are consistent with the conclusions derived from Table 1.  Specifically, all candidate Dwarf Planets with a radius exceeding 450 km have a mass exceeding $5 \times 10^{20}$ kg whereas those with a radius less than 450 km all have a mass less than $5 \times 10^{20}$ kg (Table 2).  The Trans-Neptunian bodies Orcus and Salacia are boundary cases that demonstrate the usefulness of a 450 km radius and a $5.0 \times 10^{20}$ kg  mass as the lower limits for bodies meeting the roundness criteria.  Orcus, with a radius of 455 km, has a mass slightly above the minimum mass limit whereas Salacia has a radius of 423 km and a mass slightly below the minimum mass limit (Table 2).  These two bodies illustrate that trans-Neptunian objects with a radius >450 km should generally be expected to have a mass >$5.0 \times 10^{20}$ kg and therefore meet the minimum mass necessary to be round and classified as Dwarf Planets.

With the addition of the Table 2 bodies to the sample in Table 1, the transitional ranges for "almost round" bodies are 160 – 450 km for radius and $3 \times 10^{19} – 5 \times 10^{20}$ kg for mass.  Bodies with mass and radius larger than the upper limits of these ranges are spheroidal bodies (Planets or Dwarf Planets) whereas those below these ranges are non-spheroidal.   Bodies within the transitional mass and radius ranges, such as Salacia, Varda, Pallas, and Vesta, are *Small Solar System Bodies*, not Dwarf Planets.

**~2.3 Comparing minimum mass from shape data to minimum mass for dynamical dominance**

It is possible for a non-spheroidal body to be dynamically dominant if the body has a very close-in orbit, or an orbit around a low mass star or brown dwarf (Soter 2006; Margot et al. 2024).  For a body with the minimum mass necessary to be spheroidal,  the maximum orbital semi-major axis values at which it can dynamically dominate an orbit are 0.09 AU, 0.35 AU, and 1.0 AU for orbits around stellar bodies with 1.0 $M_{Sun}$, 80 $M_{Jupiter}$, and 13 $M_{Jupiter}$ respectively (Margot et al. 2024, Figure 7).  Since numerous exoplanets have been discovered with orbital semi-major axes <0.09 AU, the minimum mass for roundness identified in this analysis is useful to differentiate "Small Stellar System Bodies" that orbit close enough to their star or brown dwarf to be dynamically dominant, but not massive enough to be spheroidal, from Planets.

**Table 2:  Dwarf Planet Candidates with radius 300 – 800 km**

| Object | Mass (kg) | Mass Reference | Radius (km) | Radius Reference |
|---|---|---|---|---|
| **Haumea** | $4.01 \times 10^{21}$ | Ragozzine et al. (2009) | 780 | Dunham et al. (2019) |
| **Makemake** | $3.1 \times 10^{21}$ | Parker et al. (2018) | 715 | Brown (2013a) |
| **Gonggong** | $1.75 \times 10^{21}$ | Kiss et al. (2019) | 615 | Kiss et al. (2019) |
| **Quaoar** | $1.21 \times 10^{21}$ | Braga-Ribas et al. (2025) | 545 | Kiss et al. (2024) |
| **Orcus** | $5.47 \times 10^{20}$ | Grundy et al. (2019) | 455 | Brown & Butler (2018) |
| **Salacia** | <$4.92 \times 10^{20}$ | Grundy et al. (2019) | 423 | Grundy et al. (2019) |
| **Varda** | $2.66 \times 10^{20}$ | Grundy et al. (2015) | 370 | Souami et al. (2020) |
| **2003 AZ$_{84}$** | $2.1 \times 10^{20}$ | Dias-Oliveira et al. (2017) | 336 | Dias-Oliveira et al. (2017) |
| **2002 UX$_{25}$** | $1.25 \times 10^{20}$ | Brown (2013b) | 330 | Brown&Butler (2017) |

Note: Dwarf Planet names are in green text and transitional Small Solar System Body names are in brown text.

## ~3. Extending the IAU Taxonomy to include Planetary-Scale Satellites

### ~3.1 Introducing the argument for Spheroidal Satellites as an additional dynamical class

While the IAU Planet and Dwarf Planet definitions are useful, they do not represent a complete taxonomy since these two definitions do not provide a classification for all spheroidal sub-stellar bodies in the Solar System. Specifically, there are 16 natural satellites in the Solar System that exceed the minimum mass of 5 x $10^{20}$ kg necessary to assume a spheroidal shape (Tables 3 and 4). Seven of these satellites are significantly larger in mass and radius than the largest known Dwarf Planet, Pluto. Two of these satellites, Ganymede and Titan, are larger in radius than the Planet Mercury. In this section, arguments are presented for including these 16 planetary-scale satellites as a third dynamical class for spheroidal sub-stellar bodies that distinguishes these large satellites from the numerous much smaller non-spheroidal satellites of the Solar System.

It is important to emphasize that the points made in this section are not intended as support for or against the different opinions to the suggestion that "moons are planets" (e.g. Runyon et al. 2017; Metzger et al. 2022; Margot et al. 2024). Instead, the purpose of this section is to highlight a gap in the current IAU taxonomy for spheroidal sub-stellar bodies found in the Solar System and to present a solution for this taxonomic gap that may be useful to all points of view on the "moons as planets" question.

The arguments for including spheroidal satellites as an additional dynamical class are grouped into three categories: (1) The mass of these moons in the context of scenarios for dynamical dominance; (2) The usage of these moons as analogs in comparative planetology; and (3) The formation of these moons can be modeled with the same mechanisms within a circumplanetary disk that are used to describe the formation of planets in a protoplanetary disk. In short, while they are not classified as IAU Planets, both Dwarf Planets and planetary-scale satellites are ***planetary-mass bodies***.

### ~3.2 Spheroidal Satellites have Dynamically Relevant Masses

The first argument for planetary-scale satellites as a third dynamical class is that these satellites have large enough masses, under the correct exoplanetary dynamical circumstances, to allow them to meet the IAU Planet definition requirements using the metric of Margot et al. (2024). The minimum mass necessary to dynamically dominate an orbit can be calculated from equation 8 of Margot et al. (2024):

$$m_{clear} = 0.001239\ m_{central}^{5/8}\ a_p^{9/8} \qquad (1)$$

In equation 1, $m_{clear}$ is the minimum mass required to dynamically dominate the orbit expressed in Earth masses, $m_{central}$ is the mass of the central body expressed in solar masses, and $a_p$ is the orbital semi-major axis or the orbiting body expressed in astronomical units. Equation 1 is for a clearing timescale of 10 billion years.

***It is significant to note that from equation 1 above, the six largest spheroidal satellites in the Solar System exceed the minimum mass necessary to dynamically dominate the orbit of the planet they are orbiting (Table 3 and Figure 2).*** The Moon's mass is ≈10 times larger than the minimum mass necessary to clear the Earth's orbit. All four of Jupiter's Galilean moons exceed the minimum mass necessary to clear Jupiter's orbit (Table 3). Titan's mass is approximately 40% larger than the minimum mass necessary to clear Saturn's orbit. It is important to recognize that the Earth, Jupiter, and Saturn are the bodies that have had the largest role in orbital clearing, and therefore are the dynamically dominant bodies in their orbital zones. Nonetheless, these six moons each

**Table 3: Satellites exceeding m$_{clear}$ of the Primary Planet's Orbit**

| Planet | m$_{clear}$* (kg) | Satellite | Msatellite (kg) |
|---|---|---|---|
| Earth | 0.74 x 10$^{22}$ | Moon | 7.35 x 10$^{22}$ |
| Jupiter | 4.73 x 10$^{22}$ | Ganymede | 14.82 x 10$^{22}$ |
| Jupiter | 4.73 x 10$^{22}$ | Callisto | 10.76 x 10$^{22}$ |
| Jupiter | 4.73 x 10$^{22}$ | Io | 8.93 x 10$^{22}$ |
| Jupiter | 4.73 x 10$^{22}$ | Europa | 4.80 x 10$^{22}$ |
| Saturn | 9.35 x 10$^{22}$ | Titan | 13.45 x 10$^{22}$ |

*m$_{clear}$ calculated from equation 1.

individually have sufficient mass to clear the orbital zones of the planet they are orbiting as derived from the metric of Margot et al. (2024). In fact, all six of the moons in Table 3 exceed the minimum mass necessary to clear Jupiter's orbit. This demonstrates that these moons, while not Planets, are **planetary-mass bodies**, and in the correct exoplanetary circumstances bodies with similar masses could meet the IAU Planet definition standards.

Additionally, it can be pointed out that all 16 spheroidal moons of the Solar System are massive enough to be dynamically dominant in a 0.10 AU orbit around a Solar Mass star or in a 1.0 AU orbit around a 13 Jupiter mass Brown Dwarf (Figure 2). As an example, from the metric of Margot et al. (2024), all 16 of the Solar System's spheroidal satellites would be dynamically dominant and meet the definition for "Planet" if they were the primary body in any of the orbits of the TRAPPIST-1 planets (Agol et al. 2021).

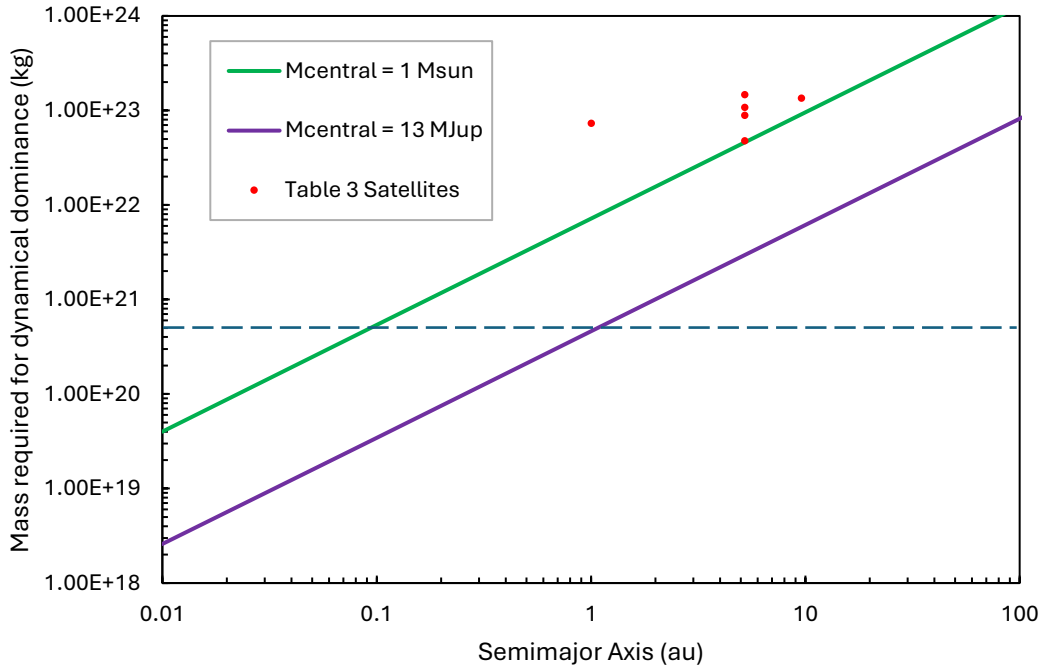

**Figure 2.** Orbital-clearing mass with increasing semimajor axis. Green line is for a 1 Solar mass star. Purple line is for a 13 Jupiter mass brown dwarf. Dashed line is the minimum mass needed to attain a spheroidal shape (5.0 x 10$^{20}$ kg). Red circles are the Satellite Planets from Table 3. Green and Purple lines calculated from Margot et al. (2024 – equation 8).

The fact that the spheroidal moons of the Solar System have sufficient mass to be dynamically dominant if substituted into realistic exoplanetary circumstances demonstrates that they are planetary-mass bodies and should be represented as a third dynamical class that distinguishes them from the numerous non-spheroidal moons of the Solar System with which they are normally grouped in the current taxonomy.

**~3.3 Spheroidal Satellites and Comparative Planetology**

The largest spheroidal moons of the Solar System have served as analogs for characterizing potential exoplanets of similar composition including exoplanet compositions that are identified as: "super-Io" (Quick et al. 2020), "super-Europa" (Vance et al. 2015), and "super-Ganymede" (Vance et al. 2015). Europa, Ganymede, and Titan are often considered analogs for cold ocean planets and water worlds (Seager et al. 2007; Vance et al. 2015; Quick et al. 2020, 2023; Kane et al. 2021).

Within the Solar System, samples for geophysical analysis and comparisons between spheroidal bodies will include the terrestrial Planets, Dwarf Planets, and spheroidal satellites. For example, Breuer et al. (2022) applied their interior modeling methods and data to Mercury, Venus, Mars, the Moon, Ganymede, and Enceladus. Stern et al. (2018) examined stagnant lid tectonics and applied their analysis to a large number of bodies in the Solar System including Planets, Dwarf Planets, and spheroidal satellites. As geoscientists, Stern et al. (2018) suggested that the terminology used by astronomers was insufficient for their purposes because their sample included not just Planets and Dwarf Planets, but also spheroidal satellites. The term "planetoids" was therefore adopted as a single general term for all large spheroidal Solar System bodies (Stern et al. 2018).

The above examples from the literature illustrate that the Solar System's large spheroidal satellites serve as important analogs to help understand planetary structure, planetary processes, and the possible characteristics of exoplanets. The usage of the spheroidal satellites in both terminology ("super-Ganymede") and geophysical comparative planetology demonstrates that these satellites, while not "Planets" dynamically, are "planetary-mass bodies" with the physical characteristics of planets.

**~3.4 Spheroidal Satellites and Planetary Formation Mechanisms**

In examining the connections between the Solar System and exoplanetary science Kane et al. (2021) noted that *"formation of regular moons, such as those in the Galilean system, may serve as analogs of compact exoplanetary systems in terms of their formation and architecture"*. Models for the formation of the Galilean moons of Jupiter include the same planet formation processes being modeled for planetary systems (see review of planet formation processes by Armitage 2024).

Models for giant satellite formation include a circumplanetary disk forming during the gas disk stage of planetary formation that is analogous to the proto-planetary disks that form planets (Canup & Ward 2002; Ward & Canup 2010; Batygin & Morbidelli 2020). Within a circumplanetary disk, successful models for satellite accretion include the same processes found in proto-planetary disks: formation of satellitesimals (Batygin & Morbidelli 2020; Madeira et al. 2021), pebble accretion (Shibaike et al. 2019; Madeira et al. 2021); streaming instability (Cilibrasi et al. 2018); and satellite core resonant chain migration (Shibaike et al. 2019; Madiera et al. 2021). The results of models for the formation of large satellites in the Solar System can be applied to exoplanetary systems (e.g. Mousis et al. 2023). Madeira et al. (2021) note that the formation of the Galilean satellites may be similar to exoplanetary systems with close-in super-Earth's such as the TRAPPIST-1 system. The architecture of the Solar System's satellite systems can also be compared with exoplanetary systems (Kane et al. 2013).

Finally, Hill (2022) demonstrated that the distribution of orbital semi-major axes for the satellites of Jupiter, Saturn, Uranus, and Neptune indicate that the large spheroidal satellites, have, since their formation, cleared their orbits of smaller satellites out to at least 5 times their Hill sphere. All satellites with radii > 500 km orbit their planet within the range $4 \times 10^5 - 4 \times 10^6$ km whereas satellites with radii <450 km all orbit closer to or farther from the primary planet than the orbital region of spheroidal satellites (Hill 2022).

### ~3.5 Summary

The examples in sections 3.2-3.4 illustrate that spheroidal satellites (1) have masses that can be dynamically relevant in the correct exoplanetary circumstances, (2) have characteristics that can serve as analogs for characterizing the composition and structure of exoplanets, (3) can form by the same processes important to the formation and architecture of planetary systems, and (4) have cleared their circumplanetary orbital zones.  All of these examples illustrate that the spheroidal satellites represent a third dynamical class for spheroidal sub-stellar bodies.  However, the IAU definitions do not include a class for spheroidal satellites which is an important gap in the IAU taxonomy.

As an additional example to illustrate this point, Neptune's satellite Triton is most likely a captured Dwarf Planet from the Kuiper Belt that shares a similar origin to Pluto (Agnor & Hamilton 2006; Nogueira et al. 2011; Bertrand et al. 2024).  As a satellite of Neptune, Triton lacks a dynamical class within the IAU taxonomy that distinguishes this captured Dwarf Planet from the numerous non-spheroidal satellites within the Solar System.  Triton, with mass and radius larger than Pluto, has sometimes been identified as a Dwarf Planet despite its absence from the IAU's official list of Dwarf Planets (Schubert et al. 2010).

The simplest solution to this  gap in the IAU taxonomy is to expand the taxonomy for spheroidal sub-stellar bodies, that began with the creation of IAU Planet and Dwarf Planet definitions, to include spheroidal satellites. Recommendations for resolving this gap in taxonomy will be described in the next section.

### ~4. Towards a More Complete Taxonomy for Spheroidal Sub-Stellar Bodies

Within the Solar System, spheroidal bodies, those with a mass exceeding the minimum mass needed to self-gravitate into a spheroid,  have three distinct dynamical circumstances.  Two of these circumstances were addressed by the IAU in 2006 with the Planet and Dwarf Planet classes.  The third dynamical class, not yet formally defined by the IAU, is represented by the 16 spheroidal satellites in the Solar System (Table 4) and any exomoons that should be confirmed (Teachey et al. 2018; Kipping 2020, 2021).   In this section, a revised taxonomy for spheroidal sub-stellar bodies is proposed.   This taxonomy incorporates concepts from the IAU 2006 resolutions and updated recommendations from Margot (2015), Lecavelier des Etangs & Lissauer (2022), and Margot et al. (2024).  The goal is to keep the taxonomic concepts, classes, terminology, and revisions already generally agreed upon, while providing additional terminology, classes, and revisions to achieve a more complete taxonomy for the classification of sub-stellar bodies.

The spheroidal sub-stellar bodies in the Solar System have three general types of orbital circumstances: (1) dynamically dominant in an orbit around the Sun, (2) not dynamically dominant in an orbit around the Sun, or (3) orbiting a larger spheroidal sub-stellar body (i.e. orbiting a Planet or Dwarf Planet). In order to capture all of these circumstances and "Small Solar System Bodies" into a single taxonomy the following definitions are suggested:

**Sub-stellar body:** A body with a mass below the minimum mass limit for core deuterium fusion (less than 13 $M_{Jupiter}$ or $2.5 \times 10^{28}$ kg).

**Planetary-mass body:** A sub-stellar body exceeding the minimum mass necessary to self-gravitate into a spheroidal shape (>$5 \times 10^{20}$ kg).

**Planet:** A planetary-mass body with a mass that exceeds the minimum mass ($m_{clear}$)[1] necessary to be dynamically dominant in its orbit around one or more stars, brown dwarfs, or stellar remnants.

**Dwarf Planet**[2]**:** A planetary-mass body with a mass below the minimum mass ($m_{clear}$) necessary to be dynamically dominant in its orbit around one or more stars, brown dwarfs, or stellar remnants.

**Satellite Planet**[2]**:** A planetary-mass body orbiting a larger mass Planet or Dwarf Planet .

**Small Solar System Body**[3]**:** A body with mass < $5.0 \times 10^{20}$ kg orbiting the Sun.

**Small Stellar System Body:** A body with mass < $5.0 \times 10^{20}$ kg orbiting a star, brown dwarf, or stellar remnant.

**satellite:** A body with mass less than $5.0 \times 10^{20}$ kg orbiting a Planet, Dwarf Planet, or larger mass Small Solar or Stellar System Body.

Notes:
~1: $m_{clear}$ is calculated with metric of Margot et al. (2024 – Equations 8 and 9).
~2: Dwarf Planets and Satellite Planets are distinct dynamical classes of bodies from Planets and are not proposed here to be considered Planets. This is consistent with the 2006 IAU resolutions but is disputed by those proposing the "geophysical planet definition" (e.g. Runyon et al. 2017, Metzger et al. 2022).
~3: Small Solar System Bodies include the various classes of asteroids, comets, Kuiper Belt objects and other Trans-Neptunian bodies.

Figure 3 shows the organization of these terms. There are a number of aspects of these suggested definitions to consider. The taxonomy retains the terms "Planet", "Dwarf Planet", and "Small Solar System Body" with the underlying conceptual meanings from the 2006 IAU resolutions. The definitions incorporate the revised language suggested that extends them to exoplanetary circumstances by including the language that Planets and Dwarf planets orbit "*one or more stars, brown dwarfs or stellar remnants*" (e.g. Margot 2015; Lecavelier des Etangs & Lissauer 2022; Margot et al. 2024) rather than "the Sun". The term "planetary-mass body" includes all sub-stellar bodies massive enough to self-gravitate into a spheroid. This single term could be used to characterize mixed samples that include Planets, Dwarf Planets, and spheroidal satellites (e.g. Stern et al. 2018). The term "Satellite Planet" is added to the taxonomy alongside the Planet and Dwarf Planet classes as the third dynamical class represented by the spheroidal planetary-mass satellites.

The term "satellite", in this taxonomy, includes all bodies with a mass less than the minimum mass necessary to be spheroidal, <$5 \times 10^{20}$ kg that are orbiting a larger mass sub-stellar body. "Satellites" are a distinct class from "Satellite Planets". Satellite Planets only orbit Planets and Dwarf Planets whereas satellites can also orbit asteroids, comets, KBOs and other trans-Neptunian bodies. The original IAU term "Small Solar System Body" is retained to represent all classes of non-spheroidal bodies orbiting the Sun. Similar bodies in exoplanetary systems can be identified as "Small Stellar System Bodies".

**Table 4: Satellite Planets of the Solar System**

| Satellite Planet | Primary Planet | Mass (kg) | Radius (km) |
|---|---|---|---|
| Ganymede | Jupiter | $1.48 \times 10^{23}$ | 2631 |
| Titan | Saturn | $1.35 \times 10^{23}$ | 2575 |
| Callisto | Jupiter | $1.08 \times 10^{23}$ | 2410 |
| Io | Jupiter | $8.93 \times 10^{22}$ | 1822 |
| Moon | Earth | $7.35 \times 10^{22}$ | 1738 |
| Europa | Jupiter | $4.80 \times 10^{22}$ | 1561 |
| Triton | Neptune | $2.10 \times 10^{22}$ | 1353 |
| Titania | Uranus | $3.46 \times 10^{21}$ | 788.9 |
| Oberon | Uranus | $3.11 \times 10^{21}$ | 761.4 |
| Rhea | Saturn | $2.31 \times 10^{21}$ | 763.8 |
| Iapetus | Saturn | $1.81 \times 10^{21}$ | 734.5 |
| Charon | Pluto | $1.59 \times 10^{21}$ | 606.0 |
| Umbriel | Uranus | $1.29 \times 10^{21}$ | 584.7 |
| Ariel | Uranus | $1.23 \times 10^{21}$ | 578.9 |
| Dione | Saturn | $1.10 \times 10^{21}$ | 561.4 |
| Tethys | Saturn | $6.175 \times 10^{20}$ | 531.1 |

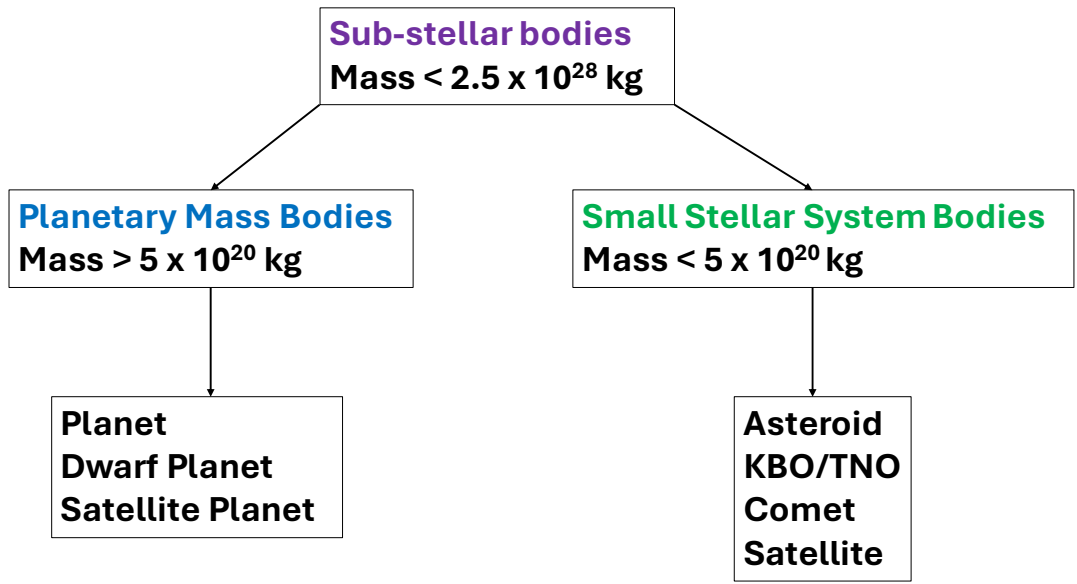

**Figure 3** – Representation of the sub-stellar taxonomy described in section 4.

## ~5. Conclusion

Triaxial ellipsoid shape data for Dwarf Planets, asteroids, and moons was examined to identify the minimum mass a body must have to self-gravitate into a "round" shape and meet the roundness criterion for the IAU Planet and Dwarf Planet definitions. The available triaxial shape data for asteroids, satellites, and Dwarf Planets in the radius range 135 – 800 km indicates that all Solar System bodies have a "round" shape, regardless of composition, if the mass exceeds $5.0 \times 10^{20}$ kg (section 2). Bodies orbiting the Sun that have a mass exceeding this mass limit are Planets or Dwarf Planets. Bodies orbiting the Sun with a mass below this mass limit are Small Solar System Bodies.

In addition to Planets and Dwarf Planets, spheroidal satellites represent a third dynamical class for spheroidal sub-stellar bodies. These planetary-mass satellites can be identified as "Satellite Planets" to distinguish them from numerous smaller non-spheroidal satellites orbiting the planets with masses $< 5.0 \times 10^{20}$ kg. There are 16 satellites in the Solar System massive enough to be identified as "Satellite Planets" (Table 4).

Arguments for identifying "Satellite Planets" as a third dynamical class include: (1) Bodies with the masses of the Solar System's spheroidal satellites are massive enough to qualify as Planets if substituted into realistic exoplanetary circumstances. For example, all 16 of the Solar System's Satellite Planets would meet the IAU Planet definition criteria if substituted into any of the orbits of the TRAPPIST-1 planets. (2) The six largest satellites in the Solar System (the Moon, Ganymede, Callisto, Io, Europa, and Titan) exceed the minimum mass necessary to dynamically dominate the orbit of the Planet they are orbiting (Table 3 and Figure 2). (3) Large spheroidal moons serve as analogs for the composition and structure of possible exoplanets both in name (i.e. super-Ganymede, super-Europa, super-Io) and in modeling their possible geophysical characteristics (e.g. cool ocean worlds). (4) The satellite formation mechanisms operating in a circumplanetary disk are the same mechanisms that form planets in a proto-planetary disk including: the formation of satellitesimals, pebble accretion, streaming instability, and migration within the disk. The formation of the Galilean moons may be analogous to the formation of close-in orbiting planets around M-dwarf stars and brown dwarfs. (5) The spheroidal satellites have all cleared their orbital zone of smaller satellites, which orbit closer to or farther from the planet that the spheroidal satellite zone (Hill 2022).

Revisions and additions to the IAU Planet and Dwarf Planet taxonomy are suggested. The changes and additions proposed here keep the concepts, terms, and revisions previously suggested (Lecavelier des Etangs & Lissauer 2022; Margot et al. 2024), and are applicable to exoplanetary systems. Definitions are suggested for the following terms: sub-stellar body, planetary-mass body, Planet, Dwarf Planet, Satellite Planet, satellite, Small Solar System Body, and Small Stellar System Body.

In addition to being applicable to exoplanetary systems, quantitative standards are applied for determining when a body is dynamically dominant in its orbit using the metric of Margot et al. (2024) and for determining if the body is massive enough to self-gravitate into a spheroid (mass $>5.0 \times 10^{20}$ kg). These quantitative criteria provide clear standards that can be used to determine when a body is a Planet, Dwarf Planet, Satellite Planet, satellite, or Small Solar System Body.

**Acknowledgements**

This research has made use of NASA's Astrophysics Data System Bibliographic Services available at: https://ui.adsabs.harvard.edu/ and the arXiv eprint service operated by Cornell University available at: https://arxiv.org/